\documentstyle[proceedings]{crckapbk}

\begin{opening}
\title{ Evolution of the Galactic Halo and Disk}

\author{Gerard Gilmore}

\institute{Institute of Astronomy, Madingley Rd. \\ Cambridge CB3 0HA, UK}

\end{opening}
\runningtitle{Evolution of the Galactic Halo and Disk}

\begin{document}

\begin{abstract}

Correlations between stellar kinematics and chemical abundances
are fossil evidence for evolutionary connections between
Galactic structural components. Extensive stellar surveys
show that the only tolerably clear distinction between galactic
components appears in the distributions of specific angular momentum.
Here the stellar metal-poor halo and the metal-rich bulge are
indistinguishable from each other, as are the thick disk and the old
disk. Each pair is
very distinct from the other. This leads to an evolutionary model
in which the metal-poor stellar halo evolves into the inner bulge,
while the thick disk is a precursor to the thin disk. These
evolutionary sequences are distinct. The galaxy is made of two
discrete `populations', one of low and one of high angular momentum.
Some (minor?) complexity is added to this picture by the debris of
late and continuing mergers, which will be especially important in
the outer stellar halo.

\end{abstract}

\section{Introduction}

Early models of Galaxy formation and evolution envisaged an isolated
set of gas clouds whose mutual tidal torques generated net angular
momentum. These clouds cooled, coalesced, fragmented locally into
stars (in some unspecified way), and collapsed to form stellar
galaxies. This picture was essentially formulated by Stromberg (1934),
expanded to include chemical evolution by Schmidt (1959) and van den
Bergh (1962), extended and popularised by Eggen, Lynden-Bell \&
Sandage (1960), and further generalised to emphasise the possible
importance of longer lived sub-structures by Searle (1977). Extensive
discussions of these developments and the many later
observationally-based attempts to clarify and verify their essential
features may be found in Gilmore \& Carswell (1987).

All these models are essentially monolithic, in that they assume the
bulk of the present galactic baryonic mass was in place very early in
Galactic evolution, and all are concerned primarily with the Galactic
stellar halo, or Population II. The relationship between the stellar
halo and the galactic disk and bulge remained problematic. In
particular, the very large difference in specific angular momentum
between the halo and the disks implied very different collapse
factors, and possible time-scales, for formation of the halo and the
disks. The halo is moreover only a tiny part of the whole Galaxy. Both
the disks and the bulge are more massive, and more chemically evolved.

The timescale and collapse factor difficulties with these models
began to be resolved with the discovery of massive dark halos, and the
realisation that their presence changed substantially the expected
evolution of a proto-galaxy, particularly for disk formation (White \&
Rees 1977; Fall \& Efstathiou 1980). In currently popular models of
galaxy formation, very different initial conditions are assumed than
are relevant to the earlier models of galaxy evolution. It remains
probable that a picture rather similar to that outlined by Eggen,
Lynden-Bell \& Sandage provides a valid description of the inner
stellar halo of the Galaxy, so that present efforts in dating globular
clusters and the oldest field stars remain focussed on such models.
Current models of the whole of galaxy evolution suggest that very late
mergers of what were really independent galaxies, not simply local
sub-condensations in the initial proto-galaxy, are the most important
physical process affecting the formation and evolution of the outer
stellar halo, the disk and perhaps the bulge.  The most appropriate
questions for current observational test based on field stars and
chemical evolution are related to the rate, timescale and effect of
such mergers, rather than to the details of the initial core about
which the Galaxy later accreted. The essential physical point is that
galaxies are not expected to be closed systems.

In this paper we consider observational methods to isolate
representative properties of the stellar populations in the old disk,
the thick disk, and the halo, and use these to address the
evolutionary questions noted above. While the oral version of this
talk additionally summarised current evolutionary models, in the
interests of space we refer to a comprehensive recent review by Silk
\& Wyse (1993) and to papers in this volume by Wyse and by White, for
such discussions, and references to the extensive relevant literature.
Some other recent useful introductions to the literature include
Gilmore, King \& van der Kruit (1989), Gilmore, Wyse \& Kuijken
(1989), Humphreys (1993) and Majewski (1993).

The essential question of immediate interest is to determine the
evolutionary relationship, if any, between the stellar populations in
the different structural components of the Galaxy.  We attempt this
below, noting that present galaxy evolution models do not presume that
there is a simple, or even a single, evolutionary path from monolithic
initial conditions to a present-day Galaxy in which all stellar
populations are inter-related. That is, we must allow for the
possibility that the evolutionary sequence which led to today's Milky
Way contains parallel paths as well as serial links.

\section{Identification of Stellar Populations }

Gravitational potentials do not respect stellar populations. All
orbits will be found in some place in the Galaxy, such as the Solar
Neighbourhood, which had initially and have retained, or which later
acquired, the relevant orbital energy and angular momentum. This has
two significant consequences: we must sift carefully through the
contents of our neighbourhood to isolate representatives of stellar
`populations' with some specific set of properties, and we must
remember that large parts of allowed parameter space will not be
sampled locally. We illustrate these factors by isolating the
statistical properties of the Galactic thick disk from local data.

\subsection{Vertical Velocity Dispersion}

The vertical velocity dispersion of K giants and FG dwarfs (and by
implication, all stars) in the Solar Neighbourhood is correlated with
chemical abundance. Bahcall, Flynn \& Gould (1992) showed the
vertical velocity dispersion of local K giants to increase
systematically with decreasing metallicity, from $\sigma_W = 14$km/s
for stars with [Fe/H]$>0$ (`young disk'), to $\sigma_W=19$km/s for stars with
$0>$[Fe/H]$>-0.5$ (`old disk'), to $\sigma_W=46\pm2$km/s for stars
with $-0.5>$[Fe/H]$>-1.0$(`thick disk').
These vertical velocity dispersions correspond to exponential vertical
scale heights of approximately 300pc for the old disk and 1kpc for the
thick disk. Thus, to isolate a sample of stars with a substantial
thick disk contribution one should look out of the Galactic Plane.
This is nicely illustrated in  Soubiran (1993), where she uses the
disk's gravitational gradient as a thick disk selection filter.

The numerical preponderance of the old disk combined with the
relatively small kinematic difference between the old disk and the
thick disk is such that nowhere can one isolate a clean sample of
either (or any) stellar population. Rather, statistical deconvolutions of
composite distribution functions are the sharpest available tool.
These analyses do of course benefit from sample selection which
maximises the thick disk contribution to the sample, so that
inevitably large samples of faint stars, thus on average far from the
Plane, are required. Mixture estimation is an appropriate algorithm in
such cases, and has been widely applied.
Such an analysis was applied by Soubiran to deduce
that the asymmetric drift of the thick disk is some 40km/s behind that
of the old disk. This provides an additional selection criterion.

\subsection{Asymmetric Drift}

Noting that the relative fraction of thick disk to old disk stars is
increased at distance of order 1-2kpc from the Plane, and the
distinction between the overlapping distribution functions of the
disks is maximised in mean galactocentric rotation velocity, allows
one to optimise a study. Ojha {\it etal} have completed the most
recent and extensive such survey, acquiring proper motions and
photometry for a complete sample of stars in directions optimised to
determine the rotation velocity of stars in
the several galactic structural components as a function of distance
from the Galactic Plane. Applying statistical deconvolution methods to
their data, they show that the asymmetric drift of the thick disk is
constant with distance, in the range $0<z<2000$pc, at a value of $\sim
-50$km/s.  There is no detectable gradient in this value with distance
from the Plane, even though some earlier studies had suspected such a
result. This result is particularly important, as it allows a reliable
calculation of the angular momentum distributions of the several
Galactic components, which in turn are an important clue as to their
evolutionary relationships.

\subsection{Chemical Abundance Distributions}

Given the results above, we infer that a sample of stars with
Galactocentric rotation in the range $V_{rot} < 100$km/s is primarily
associated withe the stellar halo, a sample with $100 < V_{rot} <
150$km/s is primarily associated with the thick disk, especially if
selected from stars distant from the Galactic Plane, while a sample of
stars with $V_{rot}>150$km/s is predominantly old disk. Applying these
expectations, we consider the distribution of chemical abundances as a
function of $V_{rot}$. The most recent such results are provided by
surveys by Norris (1994), by Schuster, Parrao, \& Contreras Martinez
(1994), and by Gilmore, Wyse \& Jones (1995), who also provide
references to the extensive earlier surveys. These results show
the thick disk to be predominant in the metallicity range from $-0.5 >
\rm [Fe/H] > -1.0$, but with suggestive evidence for a tail in the
thick disk abundance distribution below [Fe/H]=-1, and for a tail in
the stellar halo abundance distribution above [Fe/H]=-1. This leads to
a separation between the stellar halo and the thick disk in plots of
$V_{rot}$ {\it vs} [Fe/H] which is not parallel to either axis.

\section{Correlations between Chemical Abundance and Kinematics}

The methodology by which one interprets joint kinematics and chemical
abundance information is reviewed by Gilmore, Wyse \& Kuijken (1989).
Essential observational requirements are some roughly monotonic
function of time, for which chemical abundance is appropriate, and
some hopefully monotonic function of proto-Galactic collapse (or
binding energy), for which stellar orbital eccentricity was adopted.
The important point for the present to include in such a discussion
is that current observational data do {\it not} support a simple
evolutionary transition from the stellar halo into the (primordial)
disk. The observational surveys which have established the details and
the complexity of the relationship between stellar kinematics and
stellar chemical abundance distributions have required massive efforts
from many astronomers, and form the great achievement of recent
Galactic structure studies. These efforts, in large part motivated by the
early synthesis of Eggen, Lynden-Bell \& Sandage, now allow us to
extend that synthesis to the next stage of complexity.

The overlapping distribution functions in abundance with a relatively
clear distinction in galactocentric velocity suggest specific angular
momentum as a primary criterion. The current determinations of these
distributions are shown in figure 1, from Ibata \& Gilmore (1995).

\begin{figure}
\vspace{8cm}
\caption{Figure 1: Angular momentum cumulative distribution functions
for the stellar halo (dash-dot), the bulge (solid), the thick disk
(dotted) and the old disk (dashes). The similarity between the halo
and the bulge, and that between the two disks, as well as the differences
between the two groups, are apparent, and indicate their evolutionary
relationships.  }
\end{figure}

\section{Implications for Galactic Evolution}

A correlation between stellar orbital eccentricity and metallicity has
been interpreted for many years as a indication that there is an
evolutionary sequence from radial stellar orbits populated by stars
with low chemical abundances (the stellar halo population) to circular
stellar orbits populated by stars with high chemical abundances (the
disk stellar population). The stellar population of the bulge remains
somewhat vaguely located in this sequence. The most recent
determination of this relationship is presented by Twarog \&
Anthony-Twarog (1994), whose figure 10 allows a major advance in
appreciation of Galactic evolution.
There is increasing evidence, noted the discussion above, that stars
exist locally with low abundances and low eccentricity (circular)
orbits, sometimes called the `metal-weak thick disk'. Even if there
were no such stars, however, the angular momentum distributions would
still force a re-interpretaion of the evolutionary links between the
various components of the Galaxy. It is  clear
from Figure 1 that the stars of the bulge, which have high chemical
abundances, are on high eccentricity (radial) orbits. Thus such
stars are fundamentally more closely related to the lower abundance
stars of the stellar halo than are the stars of the disks, regardless
of their abundances. Similarly, low abundance thick disk stars are
more closely related to high abundance disk stars than to the stars of
the stellar halo.

This leads to a picture of Galactic evolution in which the primordial
stellar halo evolves, roughly conserving angular momentum while
self-enriching its chemical abundances and cooling into a steeper
density distribution, into the Galactic bulge. In parallel, though
probably somewhat later, in agreement with the greater collapse factor
required to explain its high specific angular momentum, the primordial
disk self-enriched, perhaps cooling and contracting somewhat, through
what is now the thick disk into the old disk and the thin disk.
An essential prediction of such a model is that the stars of the
Galactic bulge are old. Observational tests here are complex, however,
as the central regions of the Galaxy, where the Bulge is most visible,
also host the central parts of every other Galactic component. A
deconvolution process as complex as that outlined above to isolate the
properties of the thick disk will be required.

Interaction between the low angular momentum stellar populations of
the stellar halo and bulge, and the high angular momentum poulations
of the disks may have been minimal. The Galaxy is best thought of as
two discrete components, with very different and (almost)
non-overlapping distributions of specific angular momentum, but which
overlap in age, chemical abundance, and so on.

\section{Late Mergers}

The remaining substantive issues not addressed above are the role of
late mergers, and why the thick disk is thick. It is possible that
these two questions are one, as one consequence of a significant
merger is a thickening of a stellar disk.
Direct evidence for an ongoing merger event has been discovered
recently in a study by Ibata, Gilmore \& Irwin (1994).
While investigating
the kinematic structure of the Galactic bulge, they discovered a large
phase-space structure consisting of $>100$ K giant, M giant and carbon
stars in three low Galactic latitude fields. The group has a velocity
dispersion of $<10$ kms$^{-1}$, and a mean heliocentric radial velocity of
140 kms$^{-1}$, that varies by less than 5 kms$^{-1}$ over the $8^\circ$ wide
region of sky investigated kinematically.
The colour--magnitude (CM) diagram of this region of sky was obtained
from APM scans of UKST plates; over the expected Galactic CM signature
an unexpected excess was revealed, in the form of a tight CM relation
similar to that of the SMC.  A giant branch, red horizontal clump and
horizontal branch are clearly visible. Stars belonging to the low
velocity dispersion group lie on the upper giant branch of the
unexpected CM relation.  From the magnitude of the horizontal branch,
Ibata, Gilmore \& Irwin find that the object is situated $15 \pm 2$
kpc from the Galactic centre; this value agrees well with that
obtained by direct comparison to the CMD of the SMC.
An isodensity map shows the object to be elongated (with axial ratio
$\approx 3$), spanning $>10^\circ$ on the sky in a direction
perpendicular to the Galactic plane.  It is a dwarf galaxy, the
Sagittarius dwarf, probably the thrid most massive of the Galactic
(former) satellites.

\section{Conclusions}

An interesting feature of the Sagittarius dwarf is that it
contains 3 or 4 (the data remain inconclusive) globular clusters.
This provides direct evidence that any interpretation of observational
properties for stellar components, or even stellar populations, in the
Galaxy needs to recognize that a Galaxy is not a closed system. It did
not spring forth fully formed, or evolve as a single discrete monolith.

Nonetheless, present data are interpretable in terms of a Galaxy
evolving in two more or less discrete parallel paths. Low angular
momentum material formed the stellar halo, with chemically enriched
debris cooling into the Galactic bulge. This sequence is consistent
with that outlined by Eggen, Lynden-Bell and Sandage (1962).  High
angular momentum material evolved independently, forming the disk. The
early disk either formed thick, or was thickened by the last major
galactic merger, some 10Gyr ago.  Later and continuing accretion and
mergers confuse but do not obliterate the surviving evidence for this
evolutionary path.

{}  

\section{Questions}

\noindent {\bf A. Renzini} There are a few clusters in Baade's Window
that (1) are as metal-rich as the Sun, and (2) HST C-M diagrams
indicate that they are nearly as old as the halo globulars. Does that
fit with your picture in which the Bulge does not exist, or was
recently acquired? \\
{\bf Gilmore} It is an essential feature of the picture outlined here
that at least some part of the Bulge is old, and that may be
chemically evolved. A point I failed to make sufficiently clearly in
the talk is that at least some, and possibly most, of what is now in
the central parts of the Galaxy may be unrelated to the early
evolution of the Galaxy. It may have been far away long ago.

\noindent {\bf C. Pryor}  Numerical simulations (Piateh \& Pryor, AJ
submitted) suggest that a dwarf spheroidal galaxy might survive a
pericentre equal to the current distance of the Sagittarius dwarf.
The models show that tidal destruction does not produce a lumpy
distribution. The clearest signature of destruction is a strong
velocity gradient. Since Sagittarius is more luminous than Fornax, but
the same size, it seems premature to conclude it is being destroyed by
tides. Does the kinematic data resolve this? \\
\noindent {\bf Gilmore} Present data do not adequately limit the true
size of the Sagittarius dwarf. The central surface brightness is
however some 4-5 magnitudes fainter than that of Fornax, and our
calculations provide a tidal radius of 0.5kpc, or about 10\% of its
size. Future kinematic observations and dynamical modelling are
required to determine its fate and recent past orbit reliably.

\noindent{ \bf J.-C. Pecker} I was happy to see you had found a ``young
galaxy'' (or a candidate for that) in the vicinity of our own Galaxy.
You insist upon the evolutionary pattern of galaxies (halo to bulge,
not halo to disk). It sounds fine; it shows also that the ratio
`halo:disk' is a measurement of evolutionary stage (age?) for all
galaxies of the type studied. Is there any statistical evidence of a
dependance of that with the distance (measured perhaps by the
redshift?). At least all galaxies in our vicinity are not as old as
our Galaxy is, as you have shown! \\
{\bf Gilmore} I suspect that parameters such as bulge:disk ratio,
thick disk:thin disk ratio, and more generally the Hubble type,
 are set as much by the chance
distribution of size and time of mergers and by the physics which sets
up the initial conditions of lumpiness, angular momentum
distributions, early star formation rate, and such like, as uniquely by time.
Certainly the nearest few large spirals are all different in these
parameters, while there is no reason to suspect their oldest stars
differ substantially in age.


\begin{thebibliography}{}  

\bibitem{} Bahcall, J.N., Flynn, C., \& Gould, A. 1992 ApJ 389, 234

\bibitem{}Eggen, O.J., Lynden-Bell, D. \& Sandage, A., 1962 ApJ 136, 748

\bibitem{}Fall, M. \& Efstathiou, G. 1980 MNRAS 193, 189

\bibitem{} Gilmore, G., \& Carswell, R.F. (eds) 1987 {\it The Galaxy}
(Reidel, Dordrecht)

\bibitem{}Gilmore, G., King, I.R., \& van der Kruit, P.C. 1989 {\it
The Milky Way as a Galaxy} (University Science Books, CA)

\bibitem{} Gilmore, G., Wyse, R.F.G., \& Kuijken, K. 1989 ARAA 27, 555

\bibitem{} Gilmore, G., Wyse, R.F.G., \& Jones, J.B. 1995 AJ in press

\bibitem{} Humphreys, R.M. (ed) 1993 {\it The Minnesota Lectures on the
Structure and Dynamics of the Milky Way} (ASP Conf Ser vol 39)

\bibitem{} Ibata, R.A., \& Gilmore, G. 1995 MNRAS submitted

\bibitem{} Ibata, R.A., Gilmore,G. \& Irwin, M. 1994 Nature 370, 194

\bibitem{} Majewski, S.R. (ed) 1993 {\it Galaxy Evolution: The Milky
Way Perspective} (ASP Conf Ser vol 49)

\bibitem{} Norris, J.E. 1994 preprint

\bibitem{} Ojha, D.K., Bienayme. O., Robin, A.C., \& Mohan, V. 1994
A\&A 290, 771

\bibitem{} Schmidt, M. 1959 ApJ 129, 243

\bibitem{} Schuster, W.E., Parrao, L., \& Contreras Martinez, M.E.,
1994 A\&A in press

\bibitem{} Searle, L. 1977 in {\it The Evolution of Galaxies and
Stellar Populations} eds B.M. Tinsley \& R.B. Larson (Yale Univ Press,
New Haven)

\bibitem{} Silk, J., \& Wyse, R.F.G., 1993 Physics Reports 231, 295

\bibitem{} Soubiran, C., 1993 A\&A 274, 181

\bibitem{} Stromberg, G. 1934 ApJ 79, 460

\bibitem{} Twarog, B.A. \& Anthony-Twarog, B.J. 1994 AJ 107 1371.

\bibitem{} van den Bergh, S. 1962 AJ 67, 480

\bibitem{} White, S.D.M., \& Rees, M.J. 1977 MNRAS 181, 37P

\end{thebibliography}
\end{document}